\documentclass[10pt,preprint]{aastex}

\usepackage{ulem,color}

\begin{document}

\title{Radial Velocity Variations
of Photometrically Quiet, Chromospherically Inactive {\it Kepler} 
Stars: A Link Between RV Jitter and Photometric Flicker}

\author{Fabienne~A.~Bastien\altaffilmark{1},
Keivan~G.~Stassun\altaffilmark{1,2},
Joshua~Pepper\altaffilmark{1,3},
Jason~T.~Wright\altaffilmark{4,5},
Suzanne~Aigrain\altaffilmark{6},
Gibor~Basri\altaffilmark{7},
John~A.~Johnson\altaffilmark{8},
Andrew~W.~Howard\altaffilmark{9}
Lucianne~M.~Walkowicz\altaffilmark{10}}

\altaffiltext{1}{Vanderbilt University, Physics \& Astronomy Department,
1807 Station B, Nashville, TN 37235, USA}
\altaffiltext{2}{Fisk University, Department of Physics,
1000 17th Ave. N, Nashville, TN 37208, USA}
\altaffiltext{3}{Physics Department, Lehigh University, 27 Memorial Dr. W, Bethlehem, PA 18015}
\altaffiltext{4}{Center for Exoplanets and Habitable Worlds, 525 Davey
Lab, The Pennsylvania State University, University Park, PA 16803, USA}
\altaffiltext{5}{Department of Astronomy and Astrophysics, 525 Davey Lab,
The Pennsylvania State University, University Park, PA 16803, USA}
\altaffiltext{6}{Sub-department of Astrophysics, Department of Physics,
University of Oxford, Oxford OX1 3RH, UK}
\altaffiltext{7}{Astronomy Department, University of California,
Berkeley, CA 94720}
\altaffiltext{8}{Department of Astrophysics, California Institute of
Technology, MC 249-17, Pasadena, CA 91125}
\altaffiltext{9}{Institute for Astronomy, University of Hawaii, 2680 Woodlawn Drive, Honolulu, HI 96822}
\altaffiltext{10}{Department of Astrophysical Sciences, Princeton University, 4 Ivy Lane, Princeton, NJ 08544}

\begin{abstract}
We compare stellar photometric variability, as measured from {\it Kepler}
light curves by \citet{basri11}, with measurements
of radial velocity (RV) root-mean-square (RMS) variations of all
California Planet Search overlap stars.  
We newly derive rotation periods from the Kepler light curves for all of the stars
in our study sample.
The RV variations reported herein
range from less than 4 m s$^{-1}$ to 135 m s$^{-1}$, yet the stars
all
have amplitudes of photometric variability less than 3 mmag,
reflecting the preference of the RV program for chromospherically ``quiet" stars.
Despite the small size of our sample, we find with high statistical significance that the RV RMS manifests
strongly in the Fourier power spectrum of the light curve:
stars that are noisier in RV have a greater number of frequency components in the light curve.  
We also find that spot models
of the observed light curves systematically underpredict the observed RV variations by factors of $\sim$2--1000, likely because
the low level photometric variations in our sample are driven by processes not included in simple spot models.  The stars best fit by these models tend to have simpler light curves, dominated by a single relatively high amplitude component of variability.  Finally, we demonstrate that the RV RMS behavior of our sample
can be explained in the context of the photometric variability evolutionary
diagram introduced by \citet{bastien13}.  We use this diagram to derive the 
surface gravities of the stars in our sample, revealing many of them to have 
moved off the main-sequence.  
More generally, we find that the stars with
the largest RV RMS are those that have evolved onto the ``flicker floor"
sequence in that diagram, characterized by relatively low amplitude but
highly complex photometric variations which grow as the stars evolve to
become subgiants.
\end{abstract}


\section{INTRODUCTION\label{intro}}

An outstanding problem in the detection of planets via either the transit
or radial velocity (RV) methods is noise caused by stellar magnetic
activity. Manifestations of this activity, such as star spots, convective
turbulence, and granulation, can impede, and sometimes even mimic the
signals that planets produce, particularly low-mass, Earth-like ones.  
Photometric and RV characterizations of this activity are therefore of
great importance and have been the subject of a number of studies (see,
e.g., \citet{pont11}, \citet{saar98}, and \citet{wright05}).

RV noise, or ``jitter'', is a particularly pernicious problem that has
resulted in false detections \citep{queloz01} and complicated
the confirmation of transiting planets \citep{bakos10,hartman11}.
Attempts to study the
impact of stellar processes on our ability to detect planets include
\citet{dumusque11a,dumusque11b}, who simulate the effects of granulation
and starspots, respectively, on RV measurements, and \citet{saar03}, who
semi-empirically models the effect of plages on RV observations of G
dwarfs.  Empirical studies include that of \citet{wright05} which,
employing a large sample of stars observed as part of the California and
Carnegie Planet Search program, provides a relationship linking the
magnitude of RV jitter with the $B-V$ color and absolute magnitude of a star.
A challenge to all studies so far is the finding that the RV jitter
can in some cases be ``loud" even when the star is chromospherically (and  presumably photometrically) very
``quiet", indicating that for the some stars the driver of the RV jitter
is not manifested as simple photometric variability \citep{isaacson10,wright05}.

Many studies, including those we now describe, have examined 
the relationships between RV variations and photometric variability. 
Many of the photometric manifestations of stellar activity,
however, occur at the mmag level which has been largely inaccessible to
ground-based studies.

\citet{saar98}, the classic study characterizing the relationship
between RV variations and photometric variability, found
a correlation between the weighted radial velocity dispersion
(corrected for contributions from planetary companions and the mean
internal error), the effective temperature, and the stellar rotational
velocity.  Subsequent works include the examination of correlations with
other activity indicators (see, for example, \citet{martinez10}). 
\citet{saar-donahue97} employed simple models to examine how RV measurements
are affected by starspots and convective inhomogeneities, finding that the
amplitude of RV variations is related to starspot area coverage and
rotational velocity.  They additionally see that convective
inhomogeneities manifest in line bisector variations and also depend on
rotational velocity and effective temperature.  
More recently, with the public release of Kepler data,
\citet{aigrain12} propose a
model based on distributed starspots to predict the RV jitter of a star from a
well-sampled light curve.

Now, with the advent of missions like CoRot \citep{auvergne09}, MOST
\citep{walker03} and {\it Kepler} \citep{borucki10}, which
offer sub-millimagnitude photometric precision over a long time baseline for
a multitude of stars, we may re-examine empirical relationships between
photometric variability and RV variations.  
We may now extend the photometric analysis beyond measuring
the rotation period, using different ways of characterizing the light
curves and examining how these characterizations relate to activity and RV
variability. 
Importantly, we may now examine the causes of RV jitter at very photometric variability levels, and thus hope to resolve the 
mystery of RV jitter in otherwise ``quiet" stars.

Indeed, in an initial ensemble examination of the {\it Kepler} light
curves, \citet{basri11} find a wide variety in the photometric behavior of
Sun-like stars, notably including temporally coherent but non-periodic
variability and highlighting the {\it Kepler} Mission's sensitivity not only to stellar spots
but also very low-level phenomena (see also \citet{mcquillan12}).  As part
of their work, they develop a number of tools with which to study the
photometric variability of these stars, some of which we use and describe
in what follows.  More recent studies, such as \citet{aigrain12}, are
taking advantage of present capabilities to study the impact of these
photometric variations on RV measurements.

We endeavor to compare measured RV variations with the photometric
variability observed by {\it Kepler}, whose precision has unveiled the
low-level variability of a large number of stars. 
Section~\ref{data} describes our data and observations.  
Though small, our sample is
uniquely important because it is currently the only overlap set between
the highest precision light curves and the highest precision RV measurements.

Section~\ref{results} presents the results of our statistical 
comparison between the photometric and RV variability measures. We
show that the RV jitter is not correlated with the overall amplitude of
photometric variations, as might be expected if the RV jitter is
driven by features such as spots, but is strongly correlated with the
complexity of the photometric variations as measured by the number of
significant components in the Fourier spectrum of the light curve.
We compare this finding with the predictions of the simple spot model 
of \citet{aigrain12} and the simple rotational model of \citet{saar98}.  
We find that the simple spot-based model cannot reproduce the
observed RV variations for our stars, underestimating the observed 
jitter by factors of 10--1000, except for the one star with the largest 
amplitude (presumably spot driven) photometric variations.
The simple rotational model of \citet{saar98} --- which attempts to include the 
effects of plage --- fares better, underpredicting the observed RV variations
by factors of 1.5--10.

Finally, we place these results in the context of the new
photometric evolutionary diagram introduced by \citet{bastien13},
showing that the stars with the largest RV jitter variations are
those that have evolved onto the ``flicker floor" sequence in that
diagram, which appears to mark a transition in the photometric and RV
variability characteristics of otherwise quiet stars prior to and during
their evolution as subgiants.  We assess and summarize our results 
in Section~\ref{discussion}.

\section{DATA\label{data}}

\subsection{Description of Sample}
The California Planet Search is a radial velocity planet
search campaign that, using Keck Observatory in its investigations \citep{vogt94},
has monitored some stars for 15 years \citep[cf.][]{howard10}.  
A small fraction of their target stars
lie in the {\it Kepler} field, enabling a comparison between their photometric
variability characteristics, their levels of chromospheric activity and
their RV scatter.  Spectral types, RV RMS values and {\it Kepler} IDs,
among others, are listed in Table~\ref{param}.  The chromospheric activity
indices in the table are averages of the time-series measurements of
\citet{isaacson10}.  All of the stars in the sample are inactive according to the definition of \citet{baliunas95}.

Our sample of 12 stars comprises primarily G and K stars.  
Based strictly on surface gravities measured spectroscopically or obtained from 
the {\it Kepler} Input Catalog \citep{brown11}, most of these stars are dwarfs.  
However, the true $\log g$ values of the stars as determined
spectroscopically and through the analysis presented in this study
indicate that the stars span a range of $\log g$ from 2.5 to 4.5, and
thus include 7 main sequence dwarfs ($\log g \ge 4.1$), 
3 subgiants ($3.5 < \log g < 4.1$), and 2 red giants ($\log g < 3.5$). 
The available spectroscopic $\log g$ for the sample are included in 
Table~\ref{param}, as well as the $\log g$ determined from our analysis of 
the light curve ``flicker" as described in Section~\ref{sec:flicker}.

Given that the stars were selected for
the RV planet survey based on their low S index activity
\citep{duncan91}, we expect these
chromospherically inactive stars to display low levels of photometric
variability, and indeed the {\it Kepler} light curves bear this
expectation out through their very low photometric
amplitudes of less than 3 mmag (Table~\ref{param} and Fig.~\ref{lc}).
Despite the very low photometric variability amplitudes for
the entire sample, the stars exhibit a large range of RV variations;
one star, a rapidly rotating F star, has an RV RMS of 135~m~s$^{-1}$
(see Table~\ref{param}).

\subsection{Measurement of Radial Velocity Jitter\label{measrv}}

\begin{figure}[ht]
\centering
\includegraphics[scale=0.6]{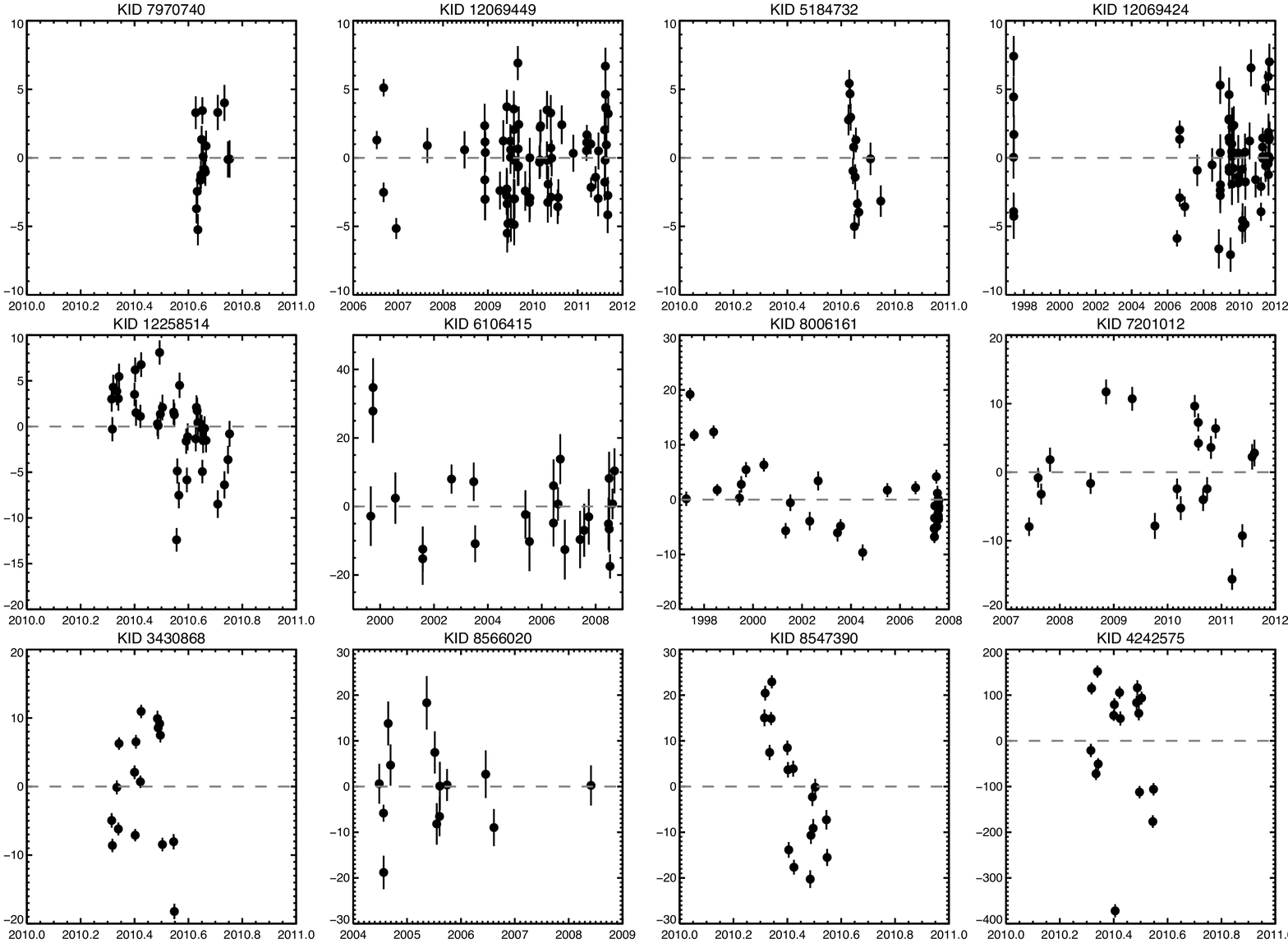}
\caption{\label{rvs}}
Radial velocity time-series of the stellar sample, with date on the x-axis.  
The y-axis is in~m~s$^{-1}$.  Note the different axis scales for each panel.  
Each plot lists the
{\it Kepler} ID in its title, and we show the stars in order of increasing RV RMS.  
Notes on individual stars are in Section~\ref{measrv}.
\end{figure}

The radial velocity time-series measurements, shown in Fig.~\ref{rvs}, were
obtained in support of the California Planet Search program and made at
the Keck and Lick Observatories (see also \citet{isaacson10}, \citet{johnson07}, and
\citet{wright05}).  For each star, we combined the velocities
into 2~hour bins, weighting each velocity by the random (``internal'')
errors derived from the variance in the velocities reported from each
section of the spectrum \citep{butler96}.  
We then calculated
the standard deviation of these binned observations' precise differential
Doppler velocities.  
We note that resulting RV RMS values less than
$\sim4$ m s$^{-1}$ may be dominated by instrumental systematics
and shot noise.  Such values reported herein are therefore upper limits.
Finally, throughout the text we use the term `RV jitter' to be synonymous with RV RMS.
We emphasize that our use of these terms does not assume nor is intended to imply
that the observed variations are simple Gaussian or `white' noise; the photometric
and RV variations we utilize in this work are in most cases substantially larger than
the measurement errors and reflect real, if stochastic, variations of astrophysical origin.

\subsubsection{Notes on Individual Stars}
{\it Kepler ID 4242575:} This rapidly rotating star has the highest RV jitter in the
sample, with an RMS of 135.5~m~s$^{-1}$, measured across a span of 3 months.  An
outlying low point drives some of this; a robust RMS calculation yields a value of
119~m~s$^{-1}$, but we adopt the former value for this work.

{\it Kepler ID 6106415:} We report 25 Lick velocities taken between 1999 and 2009.  
We reject 10 measurements taken between 1998 and 2000 which have internal measurement
errors greater than 10~m~s$^{-1}$.  The median internal measurement error for this
set of observations is 7.5~m~s$^{-1}$.  The RV RMS derived from these observations is 12.6~m~s$^{-1}$, unexpectedly elevated given this star's very low photometric amplitude.  Keck observations taken in 2013 yield an RMS of 5~m~s$^{-1}$, suggesting that instrumental effects dominated the Lick measurements.  We adopt the Keck RV RMS value in what follows.

{\it Kepler ID 12069424:} We find an RV RMS of 3.2~m~s$^{-1}$ after fitting and subtracting
a strong linear trend due to the binary motion of 16~Cyg~A about
16~Cyg~B.  In this work, we adopt an upper limit of 4~m~s$^{-1}$.  Given the internal
errors, this choice of RV RMS does not make a significant difference.

{\it Kepler ID 12069449:} We report 75 measurements taken since November 2011 for this
object after subtracting a best-fit one planet model to the known planetary companion.

{\it Kepler ID 8547390:} This star shows evidence of a linear trend in the velocities, but given
the short span of the observations, this may simply be correlated astrophysical jitter.  The
RMS about the linear trend is 9~m~s$^{-1}$.

\subsection{Measurement of Rotation Periods}\label{sec:periods}

\begin{figure}[ht]
\centering
\includegraphics[scale=0.6,angle=90]{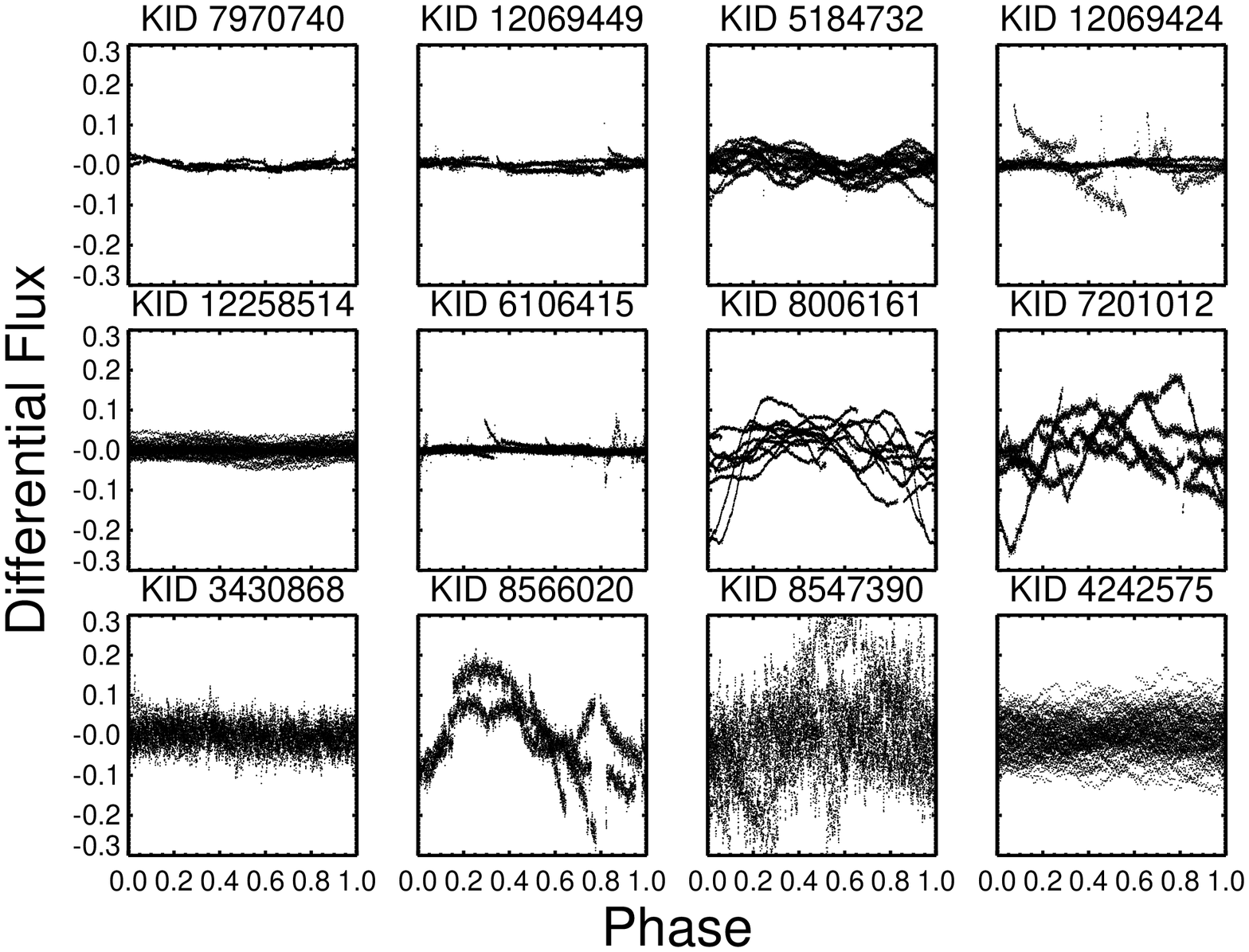}
\caption{\label{q1q4lc}}
{\it Kepler} Quarter 1 through 4 light curves, reduced as described in Section~\ref{compagrv}
and folded on their derived rotation periods.  The light curves show great complexity
on these longer timescales due, among others, to the growth and decay of active regions.  
Plots are in order of increasing RV RMS.
\end{figure}

\begin{figure}[ht]
\centering
\includegraphics[scale=0.6,angle=90]{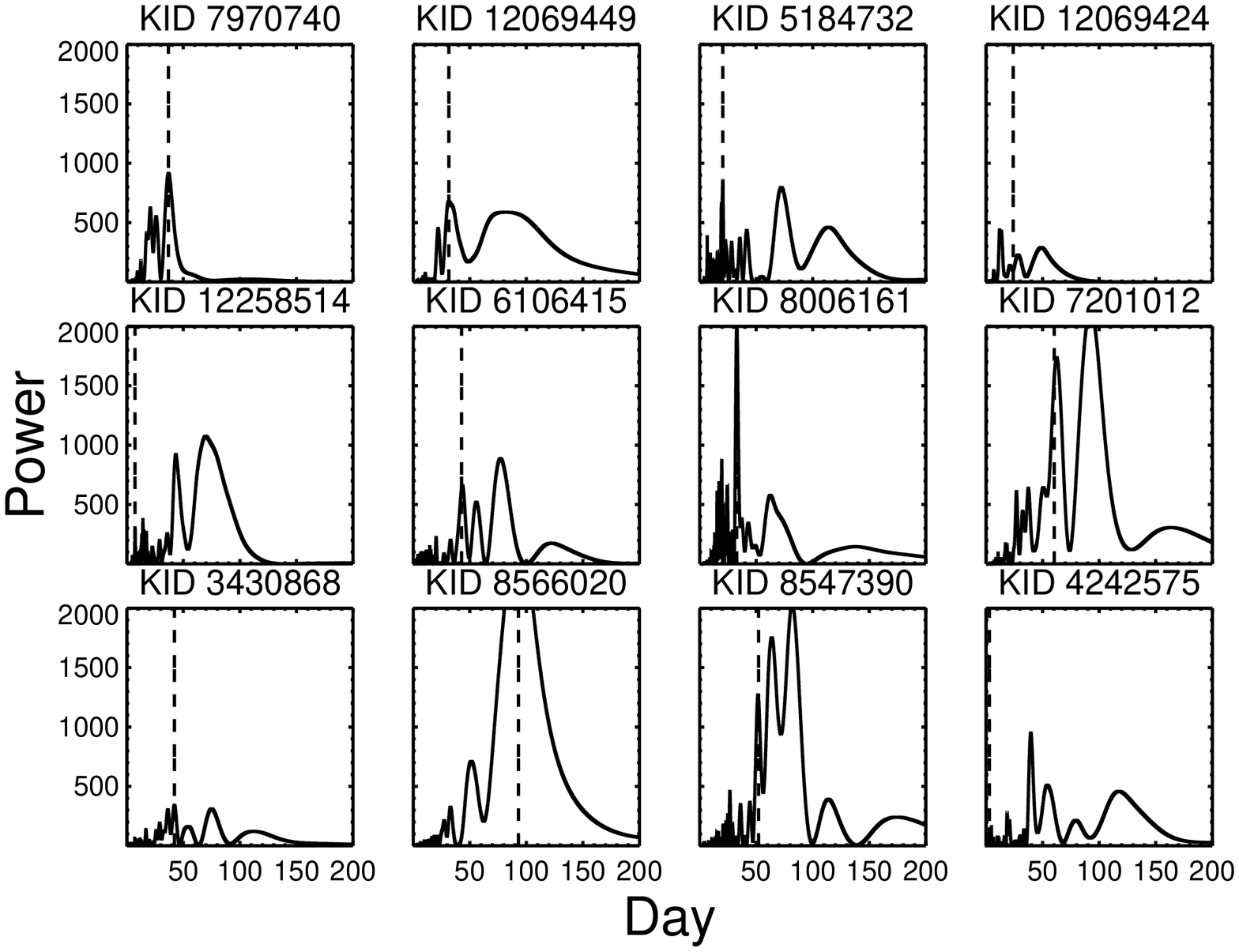}
\caption{\label{peri}}
Lomb-Scargle periodograms of the Quarters 1 through 4 {\it Kepler} light curves,
in order of increasing RV RMS.  
All plots are on the same scale.  The dashed line demarks the stellar rotation
period.  The selection of the likely rotation period is subjective due to
the complexity of the light curves and their corresponding periodograms.  See
text for notes on individual objects.
\end{figure}

We derived the periods both by visually inspecting the light
curves and by analyzing the Lomb-Scargle periodograms of those light curves \citep{lomb76,scargle82}.  
The Quarter 1--4 light curves folded on the derived periods, and the associated periodograms,
are shown in Fig.~\ref{q1q4lc}~and~\ref{peri}.
For most of the stars the periodogram reveals a clear, strong peak that we select as the likely rotation period.
In some cases (for {\it Kepler} ID 8547390, for instance), we found a few
strong contender periods (here, 52.2, 61.6 and 69.9 days) based on their power in
the Lomb-Scargle periodogram.
In such cases, the true rotation period is unclear.  
Additionally, discontinuous jumps in the reduced Quarter 3 light curve of {\it Kepler} ID 12069424,
due to instrumental artifacts, obscure the true rotation period.  We selected its
probable rotation period based on analysis of
Quarters 1, 2 and 4, together with inspection of subsequent {\it Kepler} Quarters,
rather than relying on the strongest peak in the periodogram.

\begin{figure}[ht]
\centering
\includegraphics[scale=0.6,angle=90]{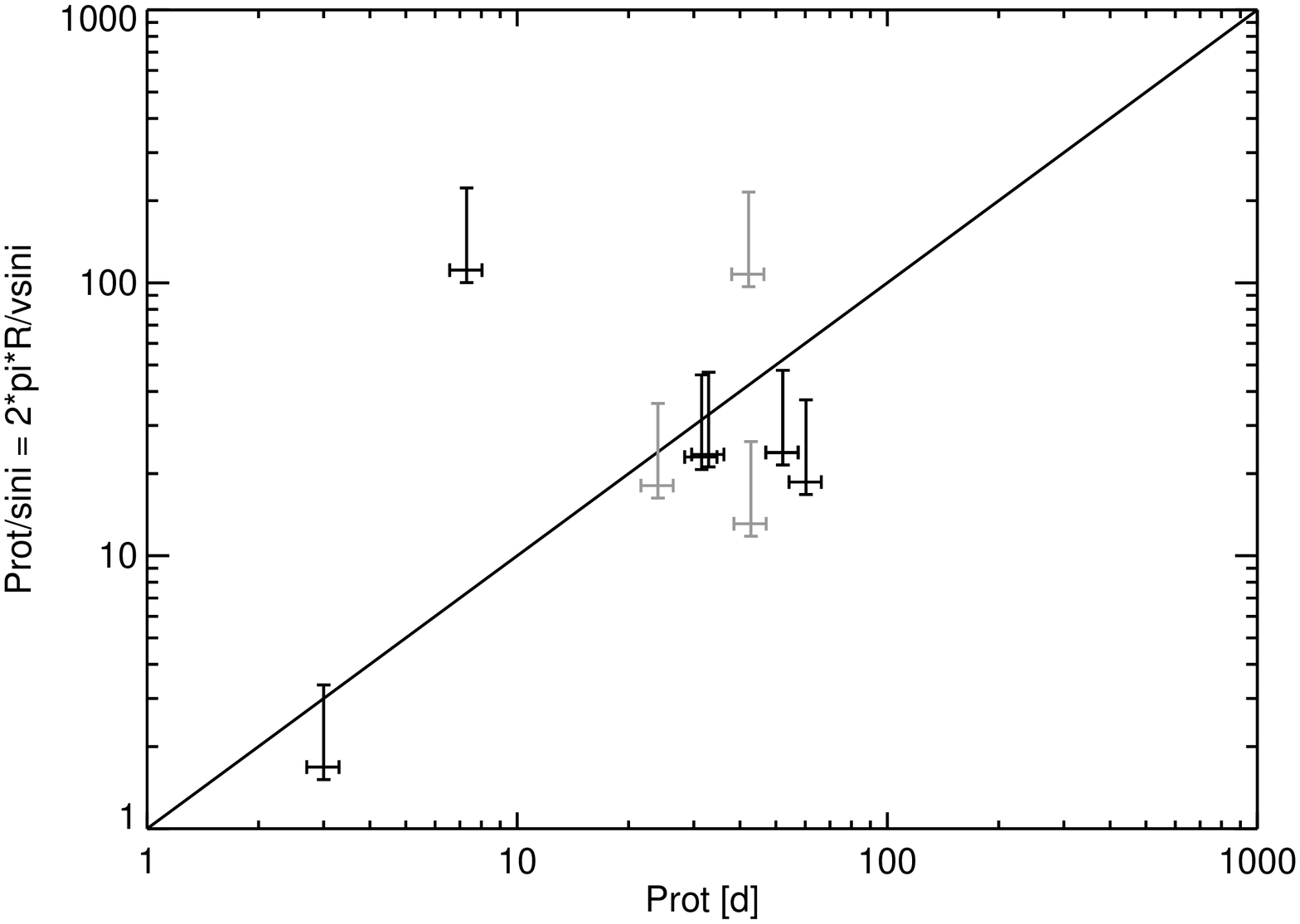}
\caption{\label{incl}}
{\bf Derived rotation periods compared with rotational velocity:}  
Where possible, we compared our photometrically measured rotation
periods with rotational velocities: the complexity of the light curves,
coupled with the low amplitudes of the photometric variations, could
otherwise result in erroneously derived rotation periods in some cases.  There nonetheless exist discrepancies, and we include quality flags on our rotation periods in Table~\ref{param}.  We represent in gray stars with a quality flag of C (signifying that the measured rotation period is unreliable).
The solid line is a
line of equality between the two plot axes.  We estimate possible 10\% errors
on the light curve rotation periods, a factor of 2 error on the $v\sin i$
in the low direction (meaning that the actual $v\sin i$ could be half the
measured value), and 10\% error on the $v\sin i$ in the high direction
(i.e., it is unlikely that the $v\sin i$ is much larger than that measured).
\end{figure}

Whenever possible, we compared our rotation periods with
$v\sin i$ measurements (Fig.~\ref{incl}), and we adopted radii from \citet{cox00}.  
In some cases the periodograms are complex and therefore our interpretation
of the periodogram for the most likely rotation period is subjective.
This complexity manifests in the folded light curves also (Fig.~\ref{q1q4lc}).
The principal results of this paper do not depend strongly on the rotation periods adopted here. Still,
the generally good correspondence between the photometrically derived rotation periods
and the periods inferred from $v\sin i$ suggests that our newly derived photometric rotation periods
are likely to be accurate.  
The rotation periods and $v\sin i$ measurements are in Table~\ref{param}.

\subsection{Photometric Variability Properties of the 
Sample\label{basri_stats}}

\citet{basri11} used {\it Kepler} Quarter 1 data to broadly characterize
the variability of all $\sim$150\,000 stars being monitored.  The
observations took place between 13 May 2009 and 15 June 2009, a span of
$\sim$33.5 days, and they restricted themselves to the Long Cadence
data, whose cadence is 29.42 minutes, for their analysis.  For each light
curve, they determined the following variability metrics:
the number of zero crossings
in the light curve smoothed by 10 hour bins (a measure of the degree of short timescale complexity in the
light curve); the variability range (a
measure of the peak-to-peak amplitude of variability in the light curve); 
from the Lomb-Scargle periodograms of the light curve, the number of
significant peaks (those whose strength is at least 10\% that of the
strongest one); and
the four-point RMS, which 
measures the
amount of high frequency variations present in the light curve 
(below we replace this with the sixteen-point RMS,
which we refer to as the ``8-hr flicker" or simply $F_8$).

\citet{basri11} additionally categorized stars 
into three groups according to both amplitude of
variability and maximal peak height in the Lomb-Scargle periodogram.  
In order from photometrically quiet to loud, the groups have
photometric amplitudes of $<$2, 2--10, and $>$10 mmag, respectively.
The quietest group in terms of the periodogram shows peaks whose 
heights are $<$30 in normalized power units.
Based on this categorization, 
all of the stars considered in this work
are photometrically very quiet, 
having variability amplitudes less than 3 mmag. 
This likely reflects the preference of the California Planet Search
target selection for the quietest stars.
But none of them meet the
{\it peak periodogram height} requirements for 
the quiet category: the maximum
periodogram heights range from $\sim38$ to $\sim715$.  {\it In other
words, the stars in this work are very low-amplitude variable stars that exhibit strong
features in their Fourier power spectra, oftentimes multiple strong features.}  
Fig.~\ref{lc} shows the light curves of the stars in this study.

\section{RESULTS\label{results}}

The principal results of this study are presented as follows.
First, we assess simple statistical correlations between RV and photometric
variability measures in order to identify the primary photometric drivers 
of RV jitter. Next, we use two different models of photomteric and RV
variations to examine the degree to which these models can reproduce the
observed correlations for photometrically quiet stars such as comprise
our sample. Finally, we place the observed photometric and RV variations
in the context of the ``flicker" evolutionary diagram newly presented by
\citet{bastien13}.

\subsection{Light Curve Periodogram 
Structure, Rather than Simple Photometric Variability, Encodes RV Jitter
\label{rvperistruct}}

We compare the measured RV RMS values with the variability
statistics developed by \citet{basri11} described in Section~\ref{basri_stats}.  To determine the
significance of
the correlations, 
we calculate a
Kendall's $\tau$ statistic, a nonparametric rank-correlation test
\citep{press92}.  Given the presence of censored data (upper limits) in
our set of RV measurements, which the canonical Kendall's $\tau$ test is not
equipped to handle, we employ the procedure of \citet{akritas96} implemented in the R statistical analysis software
package\footnote{http://www.r-project.org} (see also
\citet{helsel05}) which correctly takes censored data into account.  
We list all correlation test results for the measured RV RMS versus 
photometric variability measures in Table~\ref{lcortest}.

The amplitude of photometric variability in dwarf stars is well-known to be correlated
with the level of chromospheric activity, to the point that it is sometimes used as
a proxy for activity when no such measurement is available (as, for instance, in
\citet{chaplin11} and \citet{gilliland11}).  However, we find that RV jitter is not strongly correlated with the amplitude of photometric variability: a Kendall's
$\tau$ test yields a confidence of 80\% when
we exclude the star with the highest RV RMS and 86\% when we include it.  Such a
finding, while perhaps counterintuitive, is not surprising given that chromospheric
activity and RV RMS are only weakly correlated even in dwarfs, as \citet{saar98} demonstrate with a
sample comprised of a range of spectral types.  
We thus confirm that amplitude of photometric variability by itself is a poor
predictor of RV jitter.

\begin{figure}[ht]
\centering
\includegraphics[scale=0.6,angle=90]{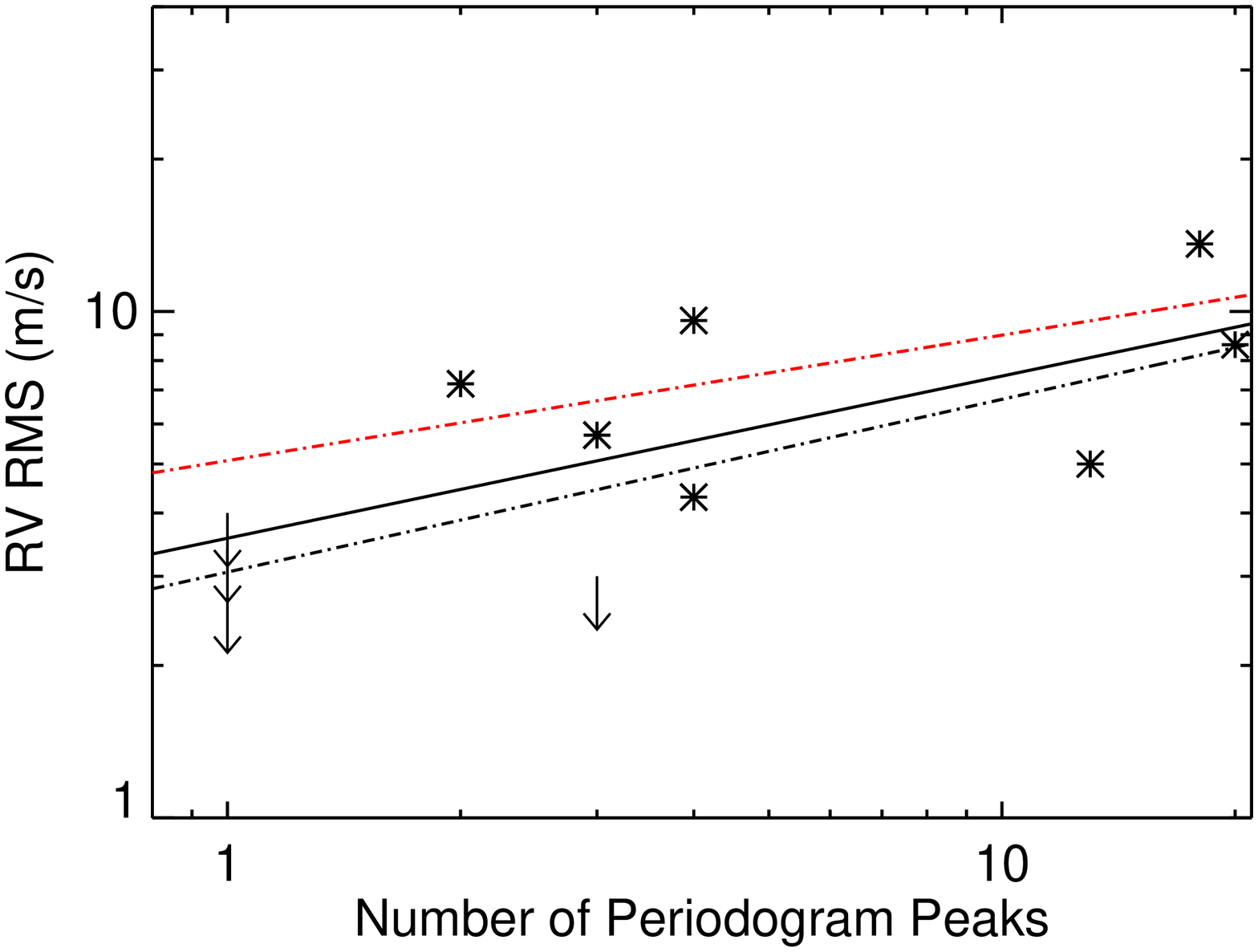}
\caption{\label{peripeaks}}
{\bf The RV RMS of low-amplitude variable stars correlates with
the number of significant periodogram peaks, }
where ``significant peaks'' are those that are at least 10\% as strong as the
highest one.  For stars with RV variations less than 20~m~s$^{-1}$ (i.e., excluding the one outlier with RV~RMS=135.5~m~s$^{-1}$), the RV RMS
correlates with the number of frequency variations in the light
curve, a finding that is statistically significant.  
An outlier, with an RV RMS of 135.5~m~s$^{-1}$, lies outside the plot.  The lines are
linear fits to the data: black lines represent linear fits to stars with RV RMS less than 20~m~s$^{-1}$.  In red is a linear fit that includes all stars in the sample.  Dot-dashed lines are more statistically robust fits to the data (see text).  The variability statistics used here were measured from the {\it Kepler} Quarter 1
light curves as reported in \citet{basri11}.
\end{figure}

Nonetheless, we find that RV jitter is strongly manifested in photometric variability in other forms,
namely in the structure of the photometric variability's Fourier power spectrum.
We show a log-log plot of the RV RMS compared with the {\it number}
of significant periodogram peaks (those at least 10\% as
strong as the maximum periodogram peak) in Fig.~\ref{peripeaks}.  This is
a key result of this work, with a
Kendall's $\tau$ confidence of 98\% (see Table~\ref{lcortest}): low-amplitude variable stars
that are noisier in RV have additional frequency components in the
corresponding light curve.  

This finding
seems to indicate that there is only one dominant frequency of variability
in the light curve of the low RV RMS stars while, as
additional significant frequencies become manifest in the light curves, the RV RMS increases.  
The Lomb-Scargle periodogram may therefore be used to estimate the RV RMS
of low-amplitude photometrically variable stars such as those examined in
this work.  We also note that this correlation holds for the 
range of evolutionary states examined here, with log(g) ranging from 
2.5 to 4.5.

Figure~\ref{peripeaks} shows a variety of linear fits to the data.  We
fit simple linear regressions to samples including and excluding the
outlier (black lines in the figure).  We additionally fit, and take as
more robust, regression lines using estimators that properly account for
the censored data points in our sample: the slope is obtained from the
Akritas-Theil-Sen estimator \citep{akritas95}, and the intercept is a
median residual from the Turnbull estimator \citep{turnbull76, helsel05}.  
We present the following as a fit to low photometric amplitude variables
with RV RMS $\lesssim$~20~m~s$^{-1}$:
\begin{equation}
RV RMS = (3.8\pm 1.7~{\rm m~s^{-1}}) \times (N_{peaks})^{0.3\pm 0.1}
\end{equation}
where $N_{\rm peaks}$ is the number of significant Lomb-Scargle periodogram
peaks, and the RV RMS is in m~s$^{-1}$.  For the uncertainties on the fit coefficients, we adopted the range of values from the three linear fits shown in Figure~\ref{peripeaks}.

\begin{figure}[ht]
\centering
\includegraphics[scale=0.6,angle=90]{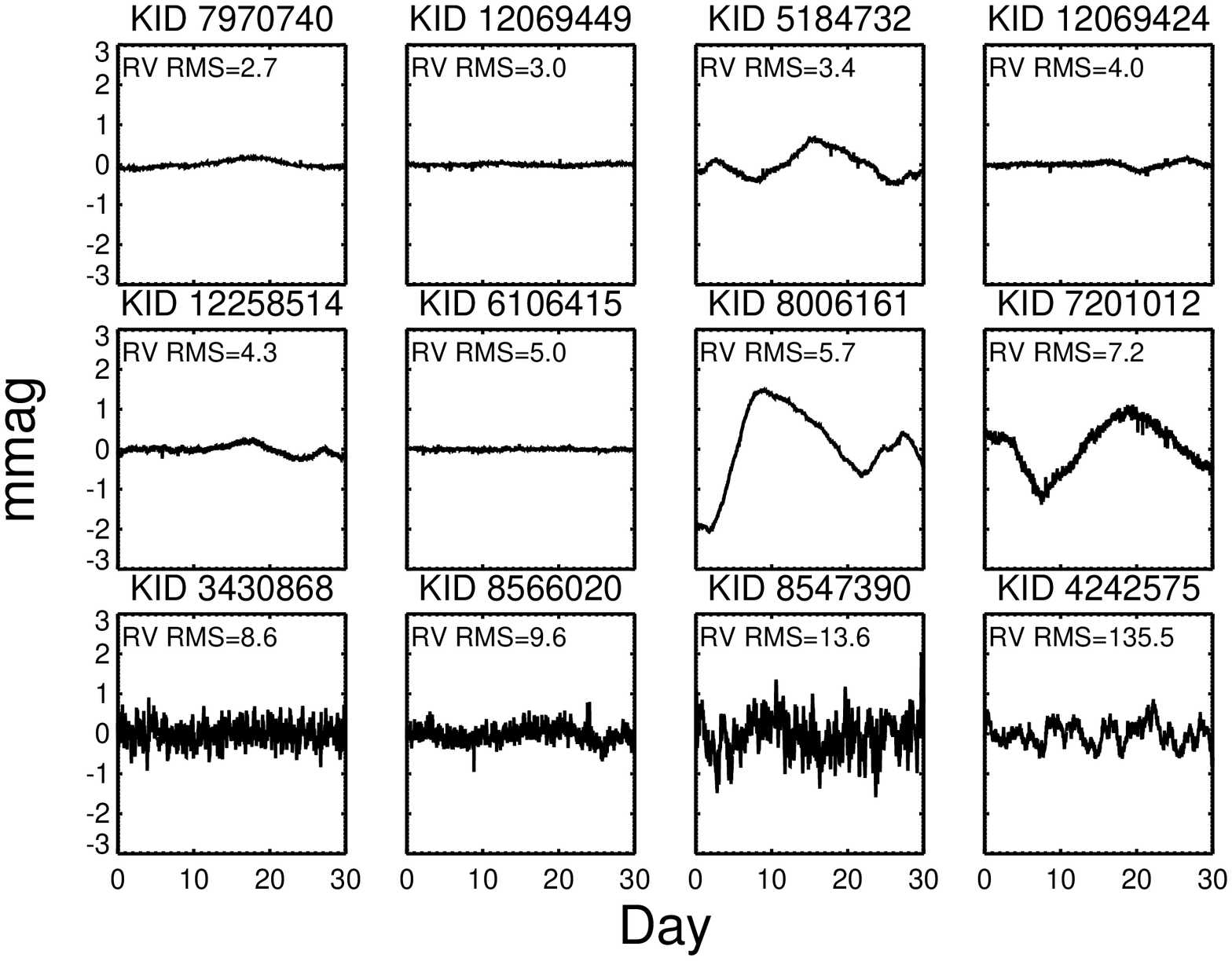}
\caption{\label{lc}}
Quarter 1 {\it Kepler} light curves of the stars examined in this study,
reduced as in \citet{basri11} and sorted in order of increasing RV RMS.  
All are shown on the same scale, and the RV RMS in m~s$^{-1}$ is indicated for each star.  
The {\it Kepler} light curves reveal that stars with higher RV RMS tend
to display higher frequency photometric variations.  
The title of each plot lists the star's {\it Kepler} ID.  We derive rotation periods from Quarters 1-4 but here only show the Quarter 1 light curve to better highlight the high frequency content of each light curve.
\end{figure}

The number of light curve zero crossings, though not highly statistically
significant on its own (Table~\ref{lcortest}), seems to corroborate this
result, providing here a measure of the high-frequency variability in the
light curves.  Indeed, the power spectra of those objects with
large $N_{\rm peaks}$ tend to show many peaks at high frequencies.  
The light curves of stars with a small number of zero crossings exhibit
longer timescale photometric variations while those light curves with
many zero crossings tend to have lower amplitudes and display higher frequency
light curve variability, evidenced by the stochasticity of their light curves
in Fig.~\ref{lc}.  

RV jitter is therefore evidently sensitive
to higher frequency photometric variability in low amplitude variable stars.  
Indeed, this trend can even be detected by visual inspection of the light curves in Fig.~\ref{lc}.
The light curves, sorted by increasing RV RMS, show a general progression in their qualitative
behavior from relatively low frequency variations to increasingly stochastic
variability. The latter stars, despite appearing ``non-variable" by traditional standards
of photometric variability amplitude, possess much more complex high-frequency variability, and in
turn display the highest levels of RV jitter.
Of course, the detailed behavior across Fig.~\ref{lc} is not simply monotonic as
some stars exhibit photometric variations with both low- and high-frequency content.
Nonetheless it is the high-frequency variability that appears to be most important for driving the RV jitter.
For example, comparing {\it Kepler} IDs 8006161 and 7201012 in Fig.~\ref{lc}, we see similar overall
amplitudes of photometric variations and similar low-frequency content. However, the latter star
also exhibits significant high-frequency variations superposed on the low-frequency variation, and
this star exhibits a larger RV RMS.  We revisit this diversity of light curve behavior and its impact on RV jitter in Section~\ref{sec:flicker}.

\subsection{Spot Models 
Systematically Under-Predict RV Jitter in Photometrically Quiet
Stars\label{compagrv} }

As noted above, the stars in our sample all have photometric 
amplitudes below $\sim$3~mmag, a poorly explored regime of Sun-like 
stars.  
We have also seen in the previous section that these stars,
despite being very photometrically quiet, do evince a strong correlation
between the RV jitter and the complexity of their light curves, as 
measured by the number of peaks in the Fourier spectra of the light curves.
Here we examine two models that have been previously developed
to estimate RV jitter from light curve variations, in order to assess
the ability of these models to reproduce the results presented above.
We find that these models systematically underestimate the observed RV 
jitter.

\subsubsection{Estimation of RV Jitter from Direct Light Curve Modelling with Spots\label{aigrain_model}}

\citet{aigrain12} provide a way to estimate RV variations due to activity
from well-sampled light curves, assuming a simple spot model, which we apply to our sample.  We processed the raw {\it Kepler} light curves to remove artifacts using a 3-point median filter with
$3\sigma$ clipping.  We then fit a straight line to the data, divided
by it, and computed the auto-correlation function of the light curve.  In
order to identify the dominant timescale of the variability, we located
the first peak where the auto-correlation function is zero, and we then
smoothed the light curve on 1/10th of that timescale using an iterative
nonlinear filter (see \citet{aigrain04}).  Finally, as in
\citet{aigrain12}, we computed the time-derivative of the light curve and
used it, along with the smoothed light curve, to simulate the RVs.  We
note that the resulting RVs should be taken with care given as the
processing of the light curves does not always properly handle
glitches and jumps in the data.  

We predicted the RV RMS with this model using the entire Quarter 1 light curve.  Figure~\ref{rvrms_div} shows the predicted versus observed RV RMS.  The predicted RV RMS is systematically too small, by factors of 10--1000 times.

Table~\ref{lcortest_ag} lists the results of comparing the
amplitudes of these simulated RVs with the various variability
statistics used in this work.  We find that the predicted RV RMS predicts a significant
correlation only with the photometric range.  
This is in strong contrast with the measured RV RMS (Table~\ref{lcortest}),
which is in fact far more sensitive to higher frequency variations than to overall range alone.  The model
used to simulate the RVs does not take such photometric variability into
account, and it therefore does not reproduce the trends we find with the
measured RV RMS.  

We find (see Fig.~\ref{rvrms_div}) that the agreement
between the measured and the simulated RVs improves when the photometric
range reaches above a threshold of $\sim$2 mmag, in effect when the photometric variations
become more simply spot dominated.  We note that the predicted RV RMS of the
rapid rotator, otherwise an outlier in this work, also has a predicted RV RMS that is significantly
discrepant with its measured RV RMS.  Its photometric amplitude is less than 1~mmag (Table~\ref{param}), highlighting the relative importance
of the amplitude of photometric variability over other factors in determining the success of such spot based models.  

The true RV measurements were not taken continuously with the high and
regular cadences of the {\it Kepler} data (which in turn become the
cadences of the predicted RV light curves).  Thus, in order to more
realistically predict the RV RMS from the light curve, we also randomly
sampled points from the calculated RV time-series, taking the same number
of data points as that used to obtain the {\it measured} RV time-series
and then calculating the predicted RV RMS.  We performed 10$^6$
realizations and then took the mean and median of the resulting
distributions.  The overall conclusions are the same as those described above.
Finally, we repeated the simulated RV RMS determination using {\it Kepler} Quarters 1--4, but the
results do not significantly differ from those obtained using the Quarter 1 light curves alone.  We thus
opted to report the results for Quarter 1 to maintain consistency with the light curve variability
statistics published by \citet{basri11}, which also used only Quarter 1 data.

It may be possible to improve the performance of this model in this regime.  For example, if our hypothesis is correct that the RV RMS is linked to short-timescale photometric variability,
then using a shorter smoothing length in the simulated RV jitter might give a better match to the measured RV jitter.

\subsubsection{Estimation of RV Jitter from a Simple Rotational Model\label{saar_model}}

We also estimate the RV jitter of the stars in our sample using the simple model 
presented by \citet{saar98}.  It is composed of two terms: a spot term and a convective term, with the latter also taking into account contributions from plage.  We use as inputs into this model the B-V colors, $v\sin i$, 
macroturbulent velocities and effective temperatures listed in Table~\ref{param}.  The model 
is also strongly rotation period dependent, and we as such we utilize the rotation periods for our sample stars as measured
in Sec.~\ref{sec:periods}.  We apply the model to our sample under the assumption that it largely contains dwarfs, as suggested by broadband photometric measurements (Table~\ref{param}).  

\begin{figure}[ht]
\centering
\includegraphics[scale=0.7, angle=90]{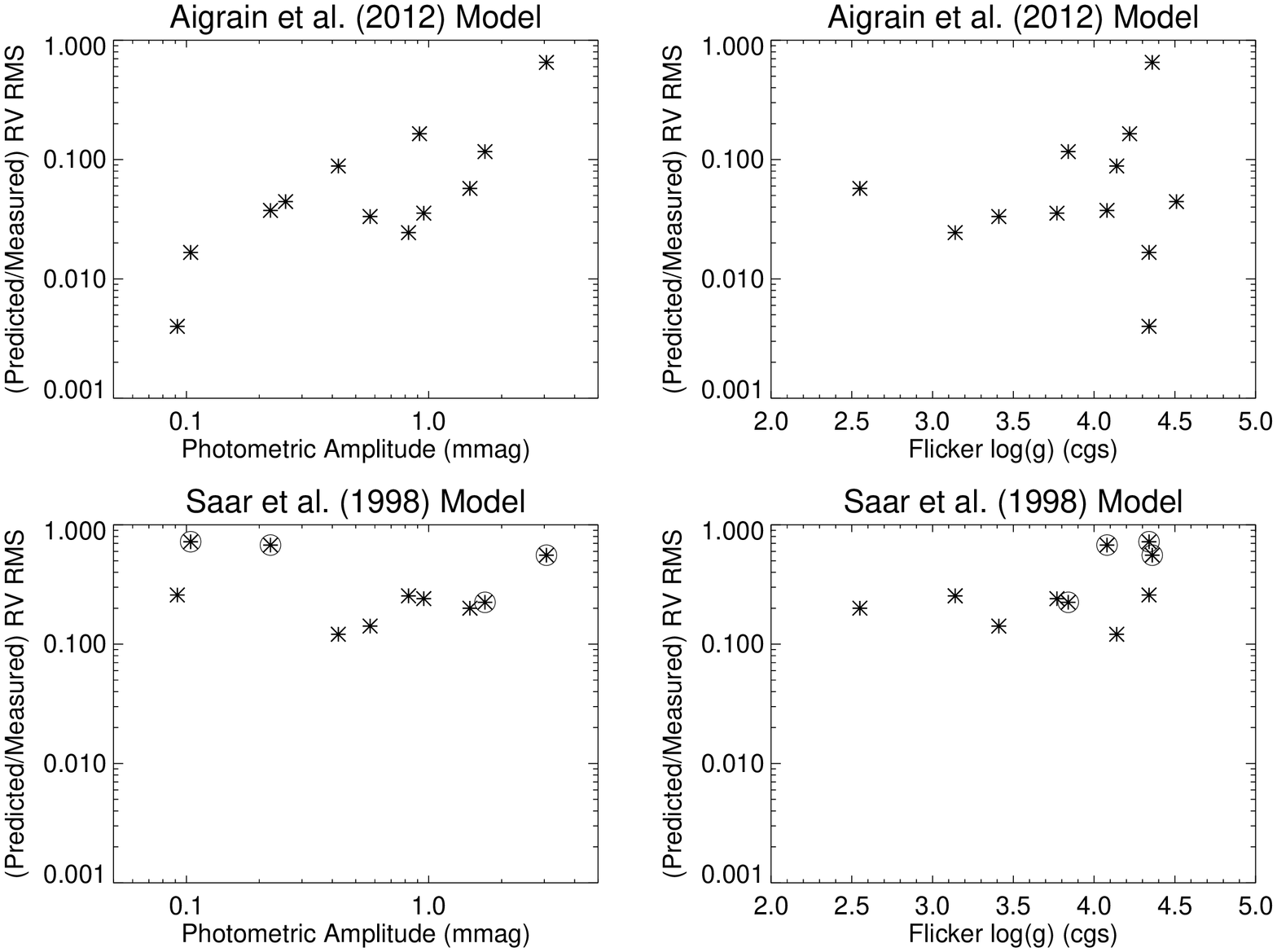}
\caption{\label{rvrms_div}}
{\bf Comparisons between measured and model-predicted RV RMS for photometrically quiet stars: }
The {\it left} panels show how the agreement between the measured RV
RMS and that predicted from two different spot models depends on the amplitude of photometric
variations.  The agreement in both cases generally improves as the photometric amplitude increases.  The photometric amplitudes shown here have not been corrected for the {\it Kepler} magnitude.
The text describes how the predicted RV RMS is derived.  The circled points in the \citet{saar98} plots are those with 
less than 3 significant light curve frequency components: the three stars which show the best agreement with model predictions have few significant Fourier components in their light curves.  {\it Right:}  similar to the above but compared against our $F_8$-based log$ g$.  There is significant spread in the agreement for log$ g$ $>$ 4. We show all plots with the same y-axis scale.
\end{figure}

As shown in 
Table~\ref{param} and Fig.~\ref{rvrms_div}, this model also systematically underestimates the actual jitter, though only by 1.5--10 times.  The agreement is better for some of the stars with gravities placing them firmly on the main sequence, as might be expected, but there are still a few instances of significant disagreement (Fig.~\ref{rvrms_div}).  The number of significant photometric variability components is a key factor in determining the success of the model: dwarfs with a smaller number of periodogram peaks tend to show better agreement with the model.  In Fig.~\ref{rvrms_div}, we highlight stars with up to three dominant Fourier components in their light curves: the three stars with \citet{saar98}-predicted values that best match observations are those with few significant Fourier components in their light curves.  Thus, limiting application of the model to true dwarfs does improve its ability to predict the RV jitter, but this by itself does not guarantee its success.  The degree of complexity in the light curve remains a crucial factor.

We note that the convective term dominates the jitter estimate for these chromospherically inactive stars, and its contribution to the RV jitter may therefore be underestimated.  One potential way to improve the convective contribution to the RV jitter is to scale this term according to the number of light curve frequency components.

\subsection{RV Jitter Correlates with Position in the Photometric 
Variability Evolutionary Diagram of \citet{bastien13} \label{sec:flicker}}

\citet{bastien13} have recently presented a new ``photometric
variability evolutionary diagram" based on analysis of a large sample of
{\it Kepler} stars. This diagram captures the time evolution of stars
from the main sequence to giant branch purely in terms of three measures
of photometric variability: Range ($R_{\rm var}$, a measure of overall
photometric variability amplitude), ``flicker" ($F_8$, a measure of
photometric variations on timescales of $<$8 hr), and
the number of light curve zero crossings ($X_0$, a measure of the light
curve complexity, similar to the number of dominant Fourier components in
the light curve discussed above).

Here we briefly summarize the salient features of this diagram from
\citet{bastien13}. First, $F_8$ was shown to be a strong correlate of
the stellar surface gravity, predicting $\log g$ with an accuracy of
$\lesssim$0.1 dex, and thus serves as a tracer of the
physical evolutionary state of a star. Stars' $F_8$ values increase systematically
with decreasing $\log g$.  Second, the diagram includes two main
populations of stars. One group consists of main sequence dwarfs with small
$F_8$ (consistent with their high $\log g$) but with a large spread of
$R_{\rm var}$, representing the spin-down evolution of main-sequence dwarfs
from rapidly rotating active stars to slowly rotating inactive stars.
Another group of stars defines a remarkably tight sequence that
\citet{bastien13} referred to as the ``flicker floor" sequence. This
sequence comprises stars with a range of
$\log g$, from dwarfs to red giants, whose $F_8$ increase as the stars
evolve. Stars on the ``flicker floor" sequence tend to be very slow
rotators, have extremely low $R_{\rm var}$ (they sit on the ``floor" of
minimum possible $R_{\rm var}$), and have as a common feature a high
degree of light curve complexity as indicated by a large $X_0$.

\begin{figure}[ht]
\includegraphics[scale=0.55]{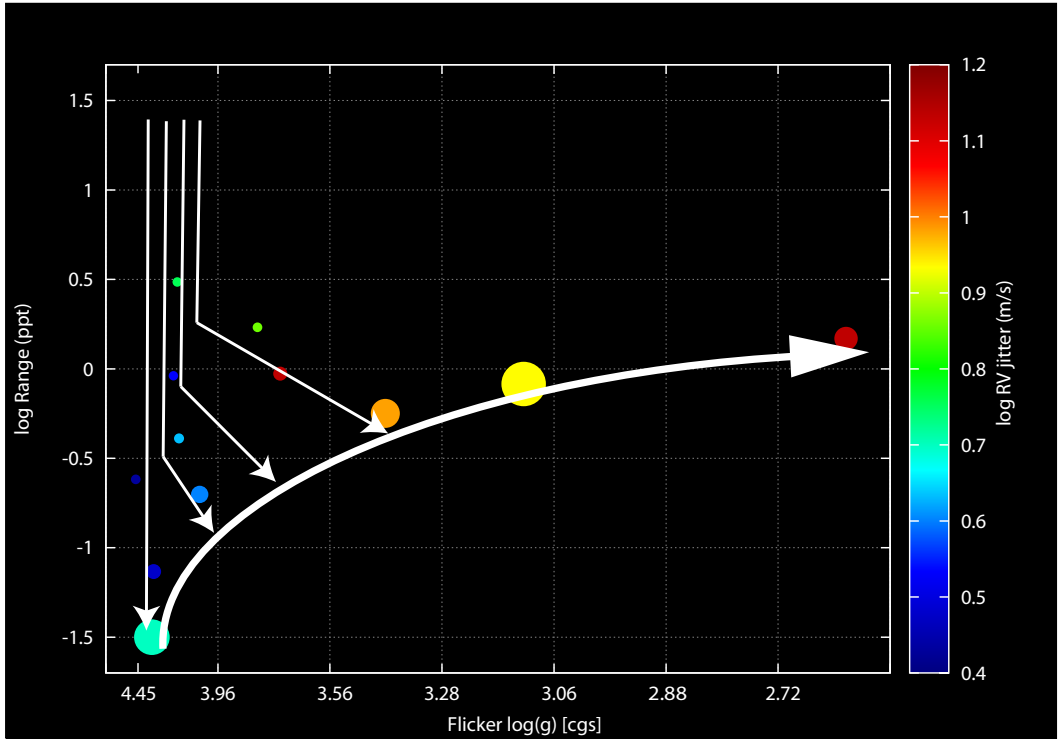}
\caption{\label{fig:bastiendiagram}
{\bf Study sample shown in the photometric variability evolutionary diagram of
\citet{bastien13}}.  Points are color-coded according to RV RMS, with the redder colors corresponding to larger RV RMS.  The arrows depict how stars evolve in this diagram (see text).  Dwarf stars are clustered to the left; those with comparatively large range show low levels of RV jitter.  Stars lying on the ``flicker floor'' tend to exhibit large levels of RV jitter.  In particular, more evolved stars show the expected higher levels of RV jitter; most of them lie on the ``flicker floor.''
}
\end{figure}

In Fig.~\ref{fig:bastiendiagram} we show our study sample in the
\citet{bastien13} photometric variability evolutionary diagram
($R_{\rm var}$ vs.\ $F_8$, with symbol size proportional to $X_0$). Here,
the RV RMS is represented by symbol color. Arrows depict the paths that
stars follow in the diagram as they evolve from rapidly rotating, active
main sequence stars at upper left, downward as they spin down and become
less active, and finally along the ``flicker floor" sequence as they
evolve as subgiants toward the red giant phase.

\begin{figure}[ht]
\includegraphics[scale=0.7, angle=90]{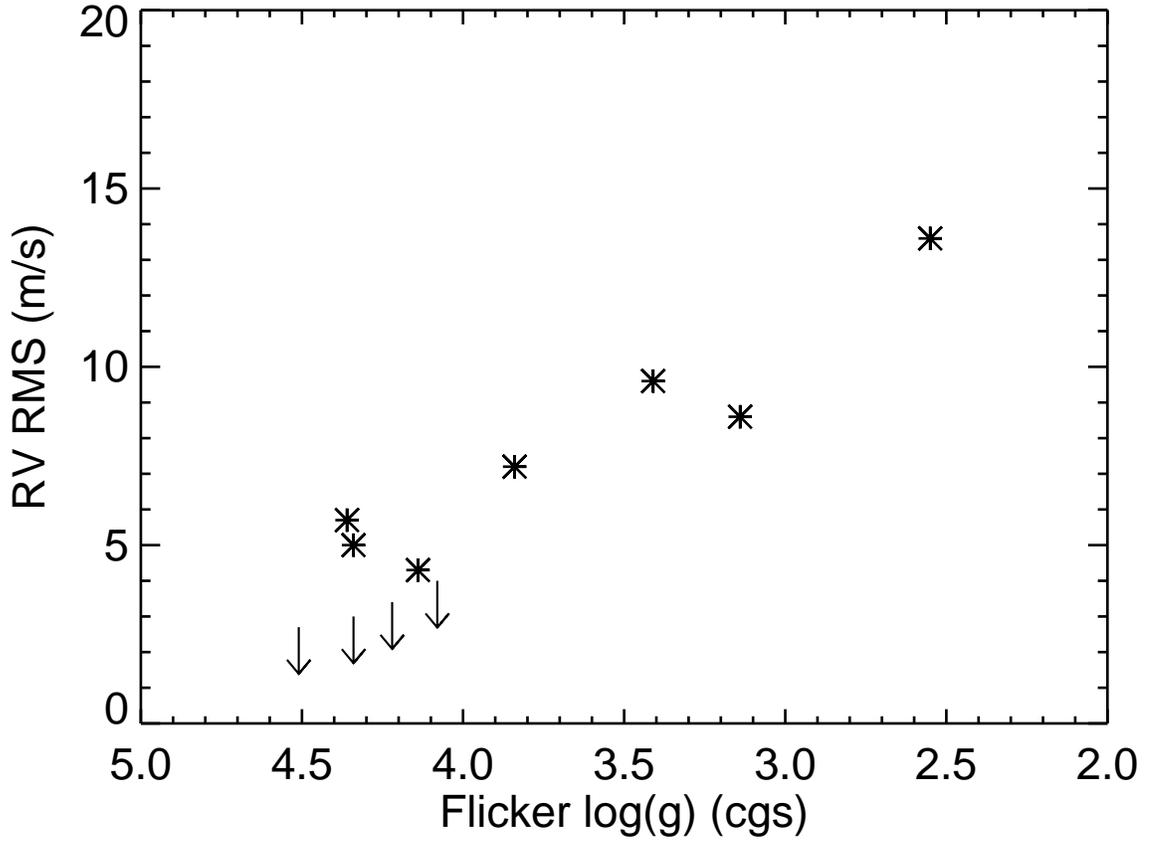}
\caption{\label{fig:rvstats}
{\bf Correlation between RV RMS and $F_8$-based log$ g$:} RV RMS shows a strong anti-correlation with $F_8$-based surface gravity, with a confidence of 97\%.  A similar trend was found by \citet{wright05}.  $F_8$  (``flicker'') measures granulation power \citep{bastien13}, indicating that the RV jitter of inactive stars is driven by convective motions on the stellar surface whose strength increases as stars evolve.
}
\end{figure}

It is clear from Fig.~\ref{fig:bastiendiagram}
that the stars with the largest RV RMS are those that have alighted
onto the flicker floor sequence. In most cases, these are stars that have
already begun their post-main-sequence evolution as subgiants, and thus
their enhanced RV jitter may be attributed to their evolved status.
The correlation between RV jitter and $F_8$ is also shown directly in
Fig.~\ref{fig:rvstats}; a Kendall's $\tau$ test gives a correlation with 97\% confidence. 

Thus, we find that the RV jitter of our sample stars can be
explained by their placement in the photometric variability evolutionary
diagram of \citet{bastien13}. In large part for our sample, which
includes several modestly evolved subgiants, the RV RMS can be predicted
simply from the stars' $F_8$ (or equivalently their $\log g$;
Fig.~\ref{fig:rvstats}).  Interestingly, there is a hint that the stellar evolutionary state is not the whole story.  In particular, one dwarf ({\it Kepler} ID 6106415) has somewhat higher RV jitter than most other dwarfs of low photometric amplitude.  What distinguishes this particular stars is its position on the $F_8$ floor, and hence its relatively large $X_0$.  More generally, the stars with the highest RV RMS, ranging from unevolved
dwarfs to evolved subgiants, are those that sit on the ``flicker floor"
sequence in the diagram.  These stars' position on the flicker floor sequence also explains the
very strong correlation between RV RMS and number of light-curve
Fourier components (Sec.~\ref{rvperistruct}) as a manifestation of these
stars' complex light curves (high $X_0$)
despite their generally very low $R_{\rm var}$.

\section{DISCUSSION AND CONCLUSIONS\label{discussion}}

In general, we find that RV RMS is quite sensitive to high-frequency
light curve variations as well as to the number of significant frequencies
that make up the light curve, apparently irrespective of spectral type for
F, G and K stars.  We find that one can use the number of significant peaks
in the Lomb-Scargle periodogram to predict the RV RMS of variable stars
whose amplitude of photometric variability is less than $\sim$3 mmag; we
provide a simple power law relation in section~\ref{rvperistruct}.  
This also manifests in the number of light curve zero crossings, a crude measure of the Fourier complexity of the light curve.
More generally, we have found compelling evidence that the stars
with the largest RV jitter are those that have alighted onto the ``flicker
floor" sequence in the photometric variability evolutionary diagram of
\citet{bastien13}, possessing low $R_{\rm var}$ but large $X_0$.

The object exhibiting the highest RV RMS is the principal outlier
throughout this work.  Though difficult to draw firm conclusions
with one data point,
we suggest that there may be different regimes of
applicability of photometric variability-RV jitter relations; the relationship between
photometric variations, particularly the number of Fourier components,
and RV RMS perhaps changes for stars with RV RMS
values exceeding $\sim$20 m~s$^{-1}$.  For such stars, the rapid rotation,
and associated high level of chromospheric activity, may suppress the high
frequency photometric variations that seem prevalent in the other stars
in our sample \citep{garcia10, chaplin11, huber11}.  

Our findings, though comprised of several modest correlations,
are based on the only sample of stars currently available with both
light curves and RV measurements of exquisite precision, permitting us to probe a regime
of activity and variability not possible heretofore. Indeed, we note
that these stars are considered inactive by most standards; see, e.g., \citet{baliunas95}.
Our findings seem to paint the following picture:
the light curve variations of the more chromospherically active stars are typically larger,
presumably dominated by the rotational modulation of simple spot regions.  
Such stars tend to have fewer dominant peaks in their Fourier spectra,
and these peaks tend to be at lower frequencies; they exhibit low levels
of RV jitter.  The opposite is true
for the slowly rotating, less chromospherically active stars:
they reside on the flicker floor sequence of the \citet{bastien13} 
photometric variability evolutionary diagram, and
the Fourier spectra of their light curves
reveal many high frequency variations (their light curves show
large $X_0$), which seem to drive their high
RV variability.  This picture only seems to hold, however, for a limited
range of RV RMS values: the applicability of RV RMS-photometric variability-
chromospheric activity relations may depend on factors such as type of
photometric variability (here we dealt with low-amplitude variables
$\lesssim$3 mmag),
the particular range of RV RMS, and a narrow range of low-activity stars.

Previous work \citep[e.g.][]{radick98} suggests that solar type stars
may undergo a transition from spot dominated activity at high activity
levels to facula-dominated activity at low activity levels, and that this transition occurs at
photometric variability levels of a few mmag. Thus our findings may suggest
that the faculae dominated variability at slow stellar rotation and low
chromospheric activity levels produces both lower amplitude photometric
variations and higher RV jitter. This interpretation would also explain
the failure of spot based models to predict the RV jitter of these faculae dominated stars.

Applying the recently developed models of \citet{aigrain12} to the {\it Kepler} light curves of our
sample stars further corroborates the above scenario. These models, which are based on using coherent spotted
stellar surfaces to model the photometric variations, accurately predict the RV jitter for the stars in our sample with
simple, single-component photometric variations, but systematically underpredict the
RV jitter for stars with multi-component, high-frequency photometric variations.

Our results hold primarily for stars with RV RMS values
below $\sim$20~m~s$^{-1}$.  In this regime, it seems that, using the
measures examined in this work, stars with few significant
Fourier components (or low $X_0$, situated well above the
flicker floor sequence in the photometric variability evolutionary diagram
of \citet{bastien13})
are better targets for RV planet searches
despite their larger amplitude photometric variations and larger chromospheric S indices.  
Conversely, those objects with larger RV variations,
while manifesting more complex light curve variations, do nonetheless in general
exhibit lower amplitude light curve variability, and thus could still serve as
good candidates for low-amplitude transit signals.

\acknowledgments
We thank Geoff Marcy and the California Planet Search
for sharing the RV RMS
values used in this work in advance of publication, and we
thank John Brewer and Debra Fischer for allowing us to report
their $v\sin i$'s, also in advance of publication.  Geoff Marcy and Howard Isaacson kindly obtained and shared with us additional Keck RV measurements of HD 177153.  We additionally
acknowledge helpful discussions with Heather Cegla, Leslie Hebb, Steve Saar and Angie Wolfgang.  
F.A.B. acknowledges support from a NASA
Harriett Jenkins pre-doctoral fellowship and from
a Vanderbilt Provost graduate fellowship.  
This work was supported in part by NASA ADAP grant NNX12AE22G to K.G.S.

The Center for Exoplanets and Habitable Worlds is supported by the
Pennsylvania State University, the Eberly College of Science, and
the Pennsylvania Space Grant Consortium.

Some of the data herein were obtained at the W.M. Keck Observatory,
which is operated as a scientific partnership among the California
Institute of Technology, the University of California and the
National Aeronautics and Space Administration.  The Observatory was
made possible by the generous financial support of the W. M. Keck
Foundation.

The authors wish to recognize and acknowledge the very significant
cultural role and reverence that the summit of Mauna Kea has always
had within the indigenous Hawaiian community.  We are most fortunate
to have the opportunity to conduct observations from this mountain.

\clearpage

\begin{deluxetable}{lrccrrrrcccccccccc}
\tablewidth{0pt}
\tabletypesize{\scriptsize}
\tablecolumns{18}
\rotate
\tablecaption{\label{param}
Stellar Parameters and Variability Statistics}
\tablehead{ \colhead{Star Name} &
            \colhead{Kepler ID} &
            \colhead{Range\tablenotemark{a}} &
            \colhead{$X_0$\tablenotemark{b}} &
            \colhead{NPP\tablenotemark{c}} &
            \colhead{P$_{rot}$\tablenotemark{d}} &
            \colhead{flag\tablenotemark{e}} &
            \colhead{$v\sin i$\tablenotemark{f}} &
            \colhead{S Index\tablenotemark{g}} & &
            \colhead{RV RMS}  & &
            \colhead{V$_{mac}$\tablenotemark{k}} & &
            \colhead{$\log g$} & &

            \colhead{B-V} &
            \colhead{T$_{eff}$} \\

            & & & & & & & & &
            \colhead{Measured\tablenotemark{h}} &
            \colhead{Aigrain\tablenotemark{i}} &
            \colhead{Saar\tablenotemark{j}} & & 
            \colhead{$F_8$\tablenotemark{l}} &
            \colhead{Spectroscopic\tablenotemark{m}} &
            \colhead{KIC\tablenotemark{n}} & &
          }
\startdata
HD 173701  &   8006161  &  3.06  &   5  &   3  &  32.9  &  A  &  1.83 &  0.214  &  5.7    &  3.72  & 3.2  & 4.6  &  4.36  & 4.53 & 3.634 & 0.843 & 5399  \\
HD 176845  &   4242575  &  0.95  &  23  &   7  &   3.0  &  B  & 33.1  &  0.208  &  135.5  &  4.82  & 32.6 & 3.2  &  3.77  &      & 4.243 & 0.528 & 6252  \\
HD 177153  &   6106415  &  0.09  &  45  &  13  &  42.8  &  C  &  4.25 &  0.154  &  5.0    &  0.02  & 1.3  & 3.6  &  4.34  & 4.20 &       & 0.569 & 5993  \\
HD 179306  &   3430868  &  0.83  &  62  &  20  &  42.2  &  C  &  0.4  &  0.115  &  8.6    &  0.21  & 2.2  & 4.7  &  3.14  &      & 4.584 & 0.910 & 5297  \\
HD 182756  &   5184732  &  0.92  &   4  &   1  &  20.5  &  A  &       &  0.147  &  3.4    &  0.56  &      &      &  4.22  &      & 4.313 &       &       \\
HD 183298  &  12258514  &  0.42  &   8  &   4  &   7.3  &  B  &  0.5  &  0.158  &  4.3    &  0.38  & 3.3  & 3.7  &  4.14  &      & 4.301 & 0.593 & 5922  \\
HD 183473  &   7201012  &  1.71  &   6  &   2  &  60.3  &  B  &  2.5  &  0.170  &  7.2    &  0.84  & 1.6  & 4.1  &  3.84  &      &       & 0.728 & 5664  \\
HD 185351  &   8566020  &  0.57  &  39  &   4  &  93.0  &  C  &  1.0  &  0.190  &  9.6    &  0.32  & 1.3  & 5.1  &  3.41  & 3.37 &       & 0.928 & 5067  \\
HD 186306  &   7970740  &  0.26  &   6  &   1  &  36.9  &  A  &       &  0.177  &  2.7    &  0.12  &      &      &  4.51  &      & 4.414 &       &       \\
HD 186408  &  12069424  &  0.22  &  25  &   1  &  24.0  &  B  &  2.8  &  0.145  &  4.0    &  0.15  & 2.7  & 4.0  &  4.08  & 4.34 &       & 0.643 & 5781  \\
HD 186427  &  12069449  &  0.10  &  21  &   3  &  31.5  &  B  &  2.2  &  0.148  &  3.0    &  0.05  & 2.2  & 4.1  &  4.34  & 4.35 &       & 0.661 & 5674  \\
HIP 93703  &   8547390  &  1.48  &  33  &  18  &  52.2  &  B  &  1.8  &  0.116  &  13.6   &  0.78  & 2.7  & 5.3  &  2.55  &      & 4.609 & 1.127 & 4919  \\

\enddata

\tablenotetext{a}{In ppt, obtained from \citet{basri11}}
\tablenotetext{b}{Number of zero crossings derived from the light curve 
smoothed by 10 hour bins \citep{basri11}}
\tablenotetext{c}{Number of significant periodogram peaks \citep{basri11}}
\tablenotetext{d}{Photometric rotation period derived from the Quarter 1--4 {\it Kepler} light curves, in days}
\tablenotetext{e}{Quality flag for rotation period. A: probable rotation period; B: questionable; 
C: improbable}
\tablenotetext{f}{In km/s}
\tablenotetext{g}{Obtained from \citet{isaacson10} and \citet{wright04}}
\tablenotetext{h}{Measured, in m/s, with planets and long term trends removed}
\tablenotetext{i}{In m/s, predicted from \citet{aigrain12}}
\tablenotetext{j}{In m/s, predicted from \citet{saar98}}
\tablenotetext{k}{Macroturbulent velocity, calculated according to \citet{valenti05}, in km/s}
\tablenotetext{l}{Derived according to \citet{bastien13}}
\tablenotetext{m}{From \citet{valenti05}; HD 185351 is unpublished}
\tablenotetext{n}{From the Kepler Input Catalog \citep{brown11}}

\end{deluxetable}

\begin{deluxetable}{lcccccc}
\tablewidth{0pt}
\tabletypesize{\scriptsize}
\tablecolumns{2}
\tablecaption{\label{lcortest}
Statistical Confidence of Correlations for Measured RV RMS Values}
\tablehead{ \colhead{Pair} &
            \colhead{Confidence\tablenotemark{a}} &
            \colhead{Confidence incl. Outlier\tablenotemark{b}} &
            \colhead{Sign of Correlation}
          }
\startdata

RV RMS vs. Variability Range                        &  80\%   &  86\%   &  positive \\
RV RMS vs. $X_0$                                    &  77\%   &  74\%   &  positive \\
RV RMS vs. Number of Periodogram Peaks              &  98\%   &  98\%   &  positive \\ 
RV RMS vs. Rotation Period                          &  97\%   &  76\%   &  positive \\
RV RMS vs. $F_8$-based log(g)                       &  97\%   &  97\%   &  negative
\enddata

\tablenotetext{a}{Statistical confidence obtained from the Kendall's 
tau statistic, properly accounting for censored data \citep{akritas96}.}

\tablenotetext{b}{Confidence obtained from the Kendall's tau statistic 
when the star with the highest RV RMS is included in the sample.}

\end{deluxetable}

\begin{deluxetable}{lcccccc}
\tablewidth{0pt}
\tabletypesize{\scriptsize}
\tablecolumns{2}
\tablecaption{\label{lcortest_ag}
Statistical Confidence of Correlations for RV RMS Values Predicted from 
Light Curve\tablenotemark{a}}
\tablehead{ \colhead{Pair} &
            \colhead{Confidence\tablenotemark{b}} &
            \colhead{Sign of Correlation}
          }
\startdata
RV RMS vs. Variability Range                        &  99.96\%  &  positive \\
RV RMS vs. $X_0$                                    &  81\%     &  negative \\
RV RMS vs. Number of Periodogram Peaks              &  17\%     &  positive \\ 
RV RMS vs. Rotation Period                          &  41\%     &  negative \\
RV RMS vs. $F_8$-based log(g)                       &  70\%     &  negative \\
\enddata

\tablenotetext{a}{Using the model of \citet{aigrain12}.  There is no apparent outlier in the RVs estimated from 
the light curve (see Table~\ref{lcortest}).}

\tablenotetext{b}{Confidence obtained from the Kendall's tau statistic.}

\end{deluxetable}

\end{document}